\documentclass[a4paper,11pt]{article}

\usepackage{pos}
\usepackage{amsmath}
\usepackage{slashed}

\renewcommand{\d}{d}
\renewcommand{\a}{\alpha}
\renewcommand{\b}{\beta}
\renewcommand{\c}{\gamma}

\newcommand{\vvarphi}{ \boldsymbol{ \varphi}}

\newcommand{\st}{
{\fontfamily{cmss}\selectfont
\text{sT}
}}
\newcommand{\tran}{
{\fontfamily{cmss}\selectfont
\text{T}
}}
\usepackage{comment}
\usepackage[export]{adjustbox}

\title{Geometrising the Micro-Cosmos on a Supermanifold}

\author[a]{Kieran Finn}
\author[a]{Viola Gattus}
\author[b]{Sotirios Karamitsos}
\author*[a,c]{Apostolos Pilaftsis}

\affiliation[a]{Department of Physics and Astronomy, University of Manchester,\\
M13 9PL, Manchester, United Kingdom}

\affiliation[b]{Dipartimento di Fisica, Universit\'a di Pisa, Largo Bruno Pontecorvo 3,\\
56127 Pisa PI, Italy}

\affiliation[c]{Theoretical Physics Department, CERN, CH-1211 Geneva 23, Switzerland}

\emailAdd{viola.gattus@manchester.ac.uk}
\emailAdd{apostolos.pilaftsis@manchester.ac.uk}
\emailAdd{sotirios.karamitsos@df.unipi.it}

\abstract{For more than half a century, covariant and differential geometric methods have been playing a central role in the development of Quantum Field Theory (QFT). After a brief historic overview of the major scientific achievements using these methods, we will focus on the covariant and differential geometric formalism originally proposed by Vilkovisky and DeWitt (VDW). We discuss recent developments made in addressing the uniqueness of the path-integral measure of the VDW effective action, and so address the problem of quantum frame dependence in cosmologically relevant scalar-tensor theories beyond the classical approximation. Particular attention will be drawn to a long-standing problem concerning the obstacles that the VDW formalism was facing from its original conception in describing generic QFTs that include fermions. We show how in addition to bosons the VDW effective action can be extended to supermanifolds to include fermions. The so-extended formulation appears to be very promising for a complete geometrisation of realistic theories of micro-cosmos, such as the Standard Model and its gravitational sector. }

\FullConference{%
  Corfu Summer Institute 2021 "School and Workshops on Elementary Particle Physics and Gravity"\\
  29 August - 9 October 2021\\
  Corfu, Greece
}

\begin{document}
\maketitle

\section{From Geometrising the Cosmos to Micro-Cosmos}

Geometry has played an instrumental role in our understanding of the cosmos in modern physics. We only have to look at General Relativity, which did away with the mysterious action at a distance for which by Newton famously ``offered no explanations", and instead explained gravitation purely as a the result of geometry of the spacetime manifold \cite{einstein}. Indeed, geometry seems to hold a certain allure, since, even in antiquity, many numerous proto-scientific or philosophical attempts were made to explain the machinery of the cosmos. Pythagorean philosophers, inspired by their idea of the Monad, viewed the universe as a unit in which all heavenly bodies move in perfect circles around the Eternal Fire, sometimes identified as the Sun itself~\cite{pythagorean}. While the Pythagorean model arose from ideas of aesthetics rather than observation, it was an early example of geometry driving the development of models of the universe. 

The principle that the cosmos could be explained through the lens of geometry was central to the debate between geocentrism and heliocentrism. The geocentric model, originally conceived by Anaximander in the 6th century BC and later advocated by Plato in the 4th century BC, was in no small part influenced by the ever-pervasive belief that humanity has a special place in the universe. However, it had to be ``patched up'' multiple times when it was found incompatible with observations, in particular retrograde motion. Secondary trajectories such as deferents and epicycles were suggested by Apollonius and Hipparcus in the 2nd century, but these were still incompatible with the non-uniform retrograde motions of the planets. Inspired further by geometry still, Ptolemy soon finally suggested a point outside of Earth, the equant,  around which the motions of the planets were uniform, finally leading to an agreement with observations~\cite{toomer1}. 

Heliocentrism, as put forward by 
Aristarchus and Seleucus in the 3rd and 2nd centuries BC, was considerably more elegant than the Ptolemaic model, but did not gain much traction, since it did not conform to Aristotelian philosophy. The latter postulated that the Earth (made as it is by the heaviest element) should sit at the centre of the cosmos, surrounded by layers of water, air, fire, and finally aether (in the form of other planets) \cite{waerden}. The Copernican revolution that occurred in the 16th century AD, sparked a wave of renewed interest in heliocentric models, something that came to be heavily frowned upon by the Church. Tycho Brahe attempted to reconcile scripture and the heliocentric model by suggesting that the Earth is orbited by the Mercury, Venus, and the Sun, which is orbited by the rest of the planets \cite{gingerich1}. Eventually, thanks to the observations of Kepler and Galileo soon after Copernicus proposed his model, heliocentrism superseded geocentrism as the dominant cosmological model, supported theoretically by Newton's law of universal attraction.
 
As the study of geometry progressed, it became clear that there was an unspoken assumption that permeated the debate between geocentrism and heliocentrism. This was the idea of an absolute \emph{frame} of reference, expressed both in geocentrism (implicitly viewing the Earth as the true centre of the Universe) and heliocentrism (Copernicus' immobile sphere of stars). The existence of an absolute frame of reference is compatible with the notion of absolute motion, as well as the existence of an immobile, permeable medium such as the luminiferous aether. However, even before the ``aether crisis'' kickstarted by the Michelson–Morley experiment \cite{morley} and the subsequent development of the theory of relativity, the notion of absolute frames was challenged on theoretical grounds by thinkers such as Leibniz, Berkeley, and Mach.  It was not long before special and general relativity made it clear that by \emph{geometrizing} the force of gravity, there is no way to elevate one frame of reference above all others.

The idea that no frame is special is now a core tenet of scientific inquiry. Indeed, this notion can be gleaned by observing that the Tychonic model and the Copernican model are equivalent; no observation can be made to distinguish them \cite{kuhn1}. Essentially, they describe the same physical setup in different frames. Despite this, a \emph{frame problem} has persisted in Quantum Field Theory (QFT) and Quantum Gravity since the 20th century. A single QFT can be described in multiple, classically equivalent ways, yet there has been considerable debate \cite{fierz, barvinsky, capozziello, faraoni, alvarez2, capozziello2, steinwachs, jarv2, kamensh, postma, domnech, jarv1, herrero, pandey, karam2, falls, nandi, francfort, karamitsos2, chiba, burns, karam} as to whether such descriptions maintain their equivalence once the theory gets quantized. As discussions about which the preferred frame should be abound in the literature, 
we take inspiration from geometry in the same way that 
lead to the development of Einstein's General Relativity, and move towards a covariant description of frames in QFT that includes fermions.

The structure of these proceedings is as follows: in Section \ref{section2} we review the issue of conformal frames in QFT and in particular the dichotomy between the so called Einstein and Jordan frames for scalar-tensor theories of gravity. We show explicitly how to write a frame-invariant metric of spacetime leading to frame invariance of the classical action. Using the frame invariant metric and field-space covariance techniques, we derive an expression for the effective action that is both frame and reparametrization invariant in Section \ref{section3}. In Section \ref{super}, we
review how to include Grassmannian coordinates in a manifold giving rise to a \textit{supermanifold}. We then examine some basic properties and operations applied to supermanifolds such as supertransposition, supertrace, etc. 
In Section \ref{section5}, we present how to employ the language of supermanifolds to define a field-space for theories with fermionic degrees of freedom. We discuss how to define field-space tensors on the supermanifold and highlight what properties should be satisfied by the field-space metric. We then describe how the metric can be obtained and calculate it explicitly for a theory with one scalar field and one Dirac fermion in Appendix \ref{appendix}. In Section \ref{section6}, we apply the covariant methods of the supermanifolds to the Vilkovisky-DeWitt (VDW) effective action so as to obtain an expression for the effective action that is fully reparametrization invariant. Our findings are then summarised in Section \ref{section7}.

\section{Conformal Frames in Field Theory}\label{section2}
Covariant methods have been used in QFT  \cite{weinberg,honerkamp,ecker,kunstatter, Burgess:1987zi,dev, huggins, lavrov, buch, franklin, cohen, sigl, finn4, finn3} and Quantum Gravity through the works of Vilkovisky and DeWitt \cite{dewitt3, dewitt2, vilk1, vilk2, dewitt1, dewitt5}. 
In particular, the VDW effective action may solve a number of problems that would require knowledge of the off-shell dynamics of a quantum system, which are highly frame dependent in the configuration space. In particular, problems of great interest and importance are as follows: (i) the gauge-independent definition of effective charges in non-Abelian gauge theories \cite{cornwall2, cornwall,pilaftsis,Aguilar:2008xm}; (ii) the gauge-invariant description of unstable particle dynamics within the context of $S$-matrix theory \cite{Papavassiliou:1995fq,Pilaftsis:1997dr,teresi}; (iii) the proper field-reparameterization invariant definition of electroweak precision observables, including Veltman’s electroweak parameter~\cite{Binosi:2009qm}; (iv) the unique field-reparameterization invariant expansions of effective field theories such as Standard Model Effective Field Theory (SMEFT) \cite{halset,Dedes:2021abc}; and (v) a frame covariant description of cosmological inflation in models with multiple scalar fields that may act as inflatons \cite{gong, moss, bounakis, fumagalli, karamitsos2, burns, karamitsos3}.

However, these methods were limited in their scope in the sense that they did not deal with the cosmological frame issue, and they did not include fermionic fields. Before we present our formalism that incorporates these desirable features, it is important to provide some background information on the topic.

A \emph{frame} in QFT (and classical theory) is simply a specific representation of the action of a theory. The two most prominent frames  are the \emph{Einstein} frame and the \emph{Jordan} frame. In the Jordan frame, there exists a \emph{non-minimal coupling} between the field and the scalar curvature, whereas in the Einstein frame, this non-minimal coupling is removed through a judicious reparametrization of the fields, including the metric. 

We will consider a scalar-tensor theory \cite{jordan1, jordan2, dicke, bergmann, wagoner, fujii, faraoni2} (that can later be quantized)  with a spin-2 graviton field $g_{\mu \nu}$ and a set of scalar fields $\varphi^A$ (collectively denoted as $\vvarphi$) with an action of the form
\begin{equation}\label{1.1}
   S^{\rm JF}[g_{\mu\nu},\boldsymbol{ \varphi}] =\int d^{D} x \sqrt{-g}\left[-\frac{f(\boldsymbol{\varphi})}{2} R+\frac{ k_{A B}}{2}(\boldsymbol{\varphi}) g^{\mu \nu}\partial_{\mu} \varphi^{A}\partial_{\nu} \varphi^{B}-V(\boldsymbol{\varphi})\right].
\end{equation}
The above action is in the Jordan frame, whereas the following action is in the Einstein frame:
\begin{equation}\label{1.2}
   S^{\rm EF}[\tilde g_{\mu\nu}, \boldsymbol{\tilde \varphi}] =\int d^{D} x \sqrt{-\tilde g}\left[-\frac{M_P^2}{2} \tilde R+\frac{1}{2} {\tilde k}_{A B}(\boldsymbol{\tilde \varphi}) {\tilde g}^{\mu \nu}\partial_{\mu} {\tilde \varphi}^{A}\partial_{\nu} {\tilde \varphi}^{B}-V(\boldsymbol{\tilde \varphi})\right],
\end{equation}
where $g\equiv \operatorname{det}(g_{\mu \nu})$ and the model functions $f(\boldsymbol{\varphi}), k_{A B}(\boldsymbol{\varphi}), V(\boldsymbol{\varphi})$ are the effective Planck mass, the scalar field-space metric and the potential respectively. These two actions are linked together by a \emph{conformal transformation} (field-dependent scaling of the metric) and a \emph{field reparametrization} (a field-depending rescaling of the fields themselves) \cite{karamitsos3}.

The classical equivalence between different frames can be seen from the observation that  $S^{\rm JF}[g_{\mu\nu},\boldsymbol{ \varphi}] =  S^{\rm EF}[\tilde g_{\mu\nu}, \boldsymbol{\tilde \varphi}]$, which indicates that the two actions are physically equivalent, in the sense that calculating their observables yields the same results. The frame problem then essentially amounts to the following question: \emph{``Upon quantization, do the Jordan- and the Einstein-frame actions yield the same observables?''} In order to probe this question at the most fundamental level, it is wise to turn to the \emph{effective action formalism}, which can be used to derive an action $\Gamma[g_{\mu\nu},\boldsymbol{\varphi}]$ that incorporates \emph{all} quantum corrections to the classical action.

Which action we use to determine the effective action may return different results. Indeed, it was found that $\Gamma^{\rm JF}[g_{\mu\nu},\boldsymbol{\varphi}] \ne \Gamma^{\rm EF}[g_{\mu\nu},\boldsymbol{\varphi}]$ except at extremal points. This is not an issue at first glance, since observable quantities are calculated at the extrema of the action. However, both for the sake of having a covariant approach in which a particular frame is not privileged above all others, and in order to ensure that off-shell formulations of QFT do not suffer from frame issues, it is certainly desirable to find a formulation which  resolves this issue.


We turn our attention to the Jordan frame scalar-tensor action \eqref{1.1}. This class of actions is described by three \emph{model functions}: the non-minimal coupling $f(\vvarphi)$, the non-canonical kinetic term $k_{AB}(\vvarphi)$, and the potential $V(\vvarphi)$. From a physical point of view, we can examine two possible transformations that essentially amount to a relabelling of variables, and should therefore not have any physical effect on the action itself: 
\begin{enumerate}
    \item Spacetime diffeomorphisms which consist of changing the coordinates on spacetime:
    \begin{equation}\label{1.31}
        x^{\mu} \rightarrow \tilde{x}^{\mu}=\tilde{x}^{\mu}\left(x^{\nu}\right),
    \end{equation}
     with invariant line elements
     \begin{equation}\label{1.4}
         d s^{2}=g_{\mu \nu} d x^{\mu} d x^{\nu}=\widetilde{g}_{\mu \nu} d \tilde{x}^{\mu} d \tilde{x}^{\nu}.
     \end{equation}
    \item  Field reparametrizations like conformal transformations and scalar field reparametrization:
    \begin{equation}
    \begin{aligned}\label{1.3}
       g_{\mu \nu} & \rightarrow \widetilde{g}_{\mu \nu}=\widetilde{g}_{\mu \nu}\left(g_{\kappa \lambda}, \boldsymbol{\varphi}\right)=\Omega^{2}(\boldsymbol{\varphi}) g_{\mu \nu}, \\
    \phi^{A} & \rightarrow \widetilde{\varphi}^{A}=\widetilde{\varphi}^{A}\left(g_{\mu \nu}, \boldsymbol{\varphi}\right)=\widetilde{\varphi}^{A}(\boldsymbol{\varphi}).
    \end{aligned} 
        \end{equation}
\end{enumerate}

The action is indeed invariant under a field reparametrization if care is taken to appropriately transform all model functions, whose transformation rules are given by \cite{karamitsos3}:
\begin{equation}\label{1.6}
  \begin{aligned}
f(\boldsymbol{\varphi})\rightarrow\tilde{f}(\boldsymbol{\varphi}) &=\Omega^{-2} f(\varphi), \\
V(\boldsymbol{\varphi}) \rightarrow \widetilde{V}(\boldsymbol{\varphi}) &=\Omega^{-4} V(\varphi),\\
k_{AB}(\boldsymbol{\varphi})\rightarrow \tilde{k}_{\widetilde{A} \widetilde{B}}(\boldsymbol{\varphi}) &=\left[k_{A B}-6 f(\ln \Omega)_{, A}(\ln \Omega)_{, B}+3 f_{, A}(\ln \Omega)_{, B}+3(\ln \Omega)_{, A} f_{, B}\right] \partial^{A} \varphi_{\widetilde{A}} \partial^{B} \varphi_{\widetilde{B}}.
\end{aligned}  
\end{equation}
However, the line element $ds^2$ is \emph{not} invariant under a conformal transformation, since it transforms as 
\begin{align}
ds^2 \to d\tilde s^2 = \Omega^2 ds^2.
\end{align}
In a gravitational theory where $g_{\mu \nu}$ is considered a dynamical field rather than the spacetime metric, this violates the invariance of the action. To rectify this, we introduce a new model function $\ell(\vvarphi)$ which transforms as \cite{finn}
\begin{equation}\label{7.3}
    \tilde{\ell}(\vvarphi) =\Omega\: \ell(\vvarphi),
\end{equation}
and use it to define a frame-invariant metric through 
\begin{equation}\label{7.2}
    {\bar g}_{\mu \nu} \equiv g_{\mu \nu}/ \ell^{2}(\vvarphi).
\end{equation}
This metric can be used to calculate a spacetime line element that is both frame- and diffeomorphism-invariant:
\begin{equation}
d \bar{s}^{2}=\bar{g}_{\mu \nu} d x^{\mu} d x^{\nu}.
\end{equation}
The strength of our approach incorporating this new function, compared to similar approaches in the literature \cite{jarv2, higgs, flanagan, catena}, is that when scaling the metric in \eqref{7.2}, no particular functional form of $\ell(\boldsymbol{\varphi})$ is specified a priori. This implies that there is no ``preferred frame", which in turn is vital in constructing an effective action that is unique and reparametrization invariant. In this approach, $\ell$ appears in the functional measure of the path integral and thus acquires the physical role of an effective Planck length, leading to complete frame invariance of the classical action S:
\begin{equation}
S\left[\tilde{g}_{\mu \nu}, \widetilde{\varphi} ; \tilde{\ell}(\varphi), \tilde{f}(\varphi), \tilde{k}(\varphi), \widetilde{V}(\varphi)\right]=S\left[g_{\mu \nu}, \varphi ; \ell(\varphi), f(\varphi), k(\varphi), V(\varphi)\right].
\end{equation}
Moreover, at tree level we have 
\begin{equation}\label{action_tree}
     S=\int d^{D} x \sqrt{-g} \mathcal{L}=\int d^{D} x \sqrt{-\bar{g}}\: \overline{\mathcal{L}}
\end{equation}
which is independent of $\ell(\boldsymbol{\varphi})$. In \eqref{action_tree} we have defined the rescaled Lagrangian
\begin{equation}
    \overline{\mathcal{L}}=\:\ell^D \mathcal{L},
\end{equation}
which is invariant under frame transformations \eqref{1.3} and spacetime
diffeomorphism~\eqref{1.31}.

\section{Covariant Methods in QFT: Bosonic Fields and Gravity}\label{section3}

So far, we have focused on the topic of conformal frames in QFT. It is now time to examine more deeply how a QFT that includes gravity as a dynamical field can be geometrised. The Vilkovisky--DeWitt approach, using the tools of differential geometry, casts the configuration space of QFT in terms of a manifold equipped with a metric and all that this implies: intrinsic curvature, parallel transport, geodesic deviation, and so on. Unlike General Relativity, this metric is determined by the model parameters of the theory; it is not dynamically determined like the spacetime metric is through Einstein's equations \cite{einstein}. 

We examine the action \eqref{1.1} with the understanding that $g_{\mu\nu}$ is a dynamical field. We therefore define an augmented field-space manifold that incorporates both the scalar fields $\phi^A$ and the gravitational tensor field $g^{\mu \nu}$ \cite{falls, fujikawa}. General coordinates in this manifold are denoted by
\begin{equation}
    \Phi^{I} = \left(\begin{array}{c}g^{\mu \nu} \\ \phi^{A}\end{array}\right), 
\end{equation}
where $I=\{\mu\nu, A\}$. Reparametrizations of the fields are then just diffeomorphisms
\begin{equation}\label{8.2}
    \Phi^I=\widetilde{\Phi}^I(\boldsymbol{\Phi})
\end{equation}
of what we call the \textit{grand field space}.

We must now define the metric in a unique way using the Lagrangian $\mathcal{L}$:
\begin{equation}\label{2.3}
    G_{A B} \equiv \frac{g_{\mu \nu}}{D} \frac{\partial^{2} \mathcal{L}}{\partial\left(\partial_{\mu} \phi^{A}\right) \partial\left(\partial_{\nu} \phi^{B}\right)},
\end{equation}
where $\mathcal{L}$ must also include the usual gauge-fixing terms.
Using \eqref{2.3} on the action \eqref{1.1} in the absence of matter fields along with the gauge-fixing terms yields the following expression for the gravitational field space metric 
\begin{equation}
G_{(\mu \nu)(\rho \sigma)}=\frac{1}{2}\left(g_{\mu \rho} g_{\sigma \nu}+g_{\mu \sigma} g_{\rho \nu}-\alpha g_{\mu \nu} g_{\rho \sigma}\right),
\end{equation}
where $\a =\a(\chi^\mu,\gamma)$ is a constant that depends on the gauge fixing condition $\chi^\mu$ and the constant~$\gamma$. The notation $(\mu \nu)$ implies that no order is specified for the spacetime indices $\mu$ and $\nu$.  To fix the constant $\a$, the chosen gauge fixing condition is 
\begin{equation}
G^{(\mu \nu)(\rho \sigma)}=g^{\alpha \mu} g^{\beta \nu} g^{\kappa \rho} g^{\lambda \sigma} G_{(\alpha \beta)(\kappa \lambda)},
\end{equation}
which in turn implies that $\a=0,1$ in four dimensions. Choosing $\a=1$ yields Vilkovisky's metric for gravity $P_{\mu \nu \rho \sigma} = G_{(\mu \nu)(\rho \sigma)}\{\a=1\}$ \cite{vilk2}.\newline



%
%
 
We now promote the field space to a configuration space. This takes into account the spacetime dependence of the fields, which means that the general set of coordinates is now given by
\begin{equation}
    \Phi^{i} \equiv \Phi^{I}\left(x_{I}\right)=\left(\begin{array}{c}g^{\mu \nu}(x) \\ \phi^{A}\left(x_{A}\right)\end{array}\right), 
\end{equation}
with $i=\left\{I, x_{I}\right\}$ and $x_{I}=\left\{x, x_{A}\right\} $.
In what follows we will employ Eistein-DeWitt notation $\phi^a \equiv \phi^A(x_A)$, and where repeated configuration space indices imply summation over discrete field space indices and integration over spacetime.

The standard definition of the effective action $\Gamma[ \boldsymbol{ \varphi}]$ can be implicitly written as follows:
\begin{align}
\label{eq:standard effective action}
 \exp\left(\frac{i}{\hbar} \Gamma[ \boldsymbol{ \varphi}]  \right)  
 = 
 \int [\mathcal{D} \boldsymbol{ \phi}]\,  \exp \left\{\frac{i}{\hbar}  \Big[ S[    \boldsymbol{ \phi}] + \frac{\delta\Gamma[ \boldsymbol{ \varphi}]}{\delta\varphi^a} (\varphi^a - \phi^a ) \Big]\right\},
\end{align}
where $\varphi^a$ denotes now the classical field. 
 The impetus for the development of the Vilkovisky--DeWitt formalism was the observation that this expression for the effective action does not act as a scalar under field reparametrizations because of the explicit presence of the fields $\varphi^a$, which are not covariant (just like $x^\mu$ is not a covariant quantity in GR). 

Incorporating our model parameter $\ell(\vvarphi)$, we modify the functional derivative such that it takes into account conformal transformations through the definition of the frame-invariant (in addition to diffeomorphism-invariant as usual) delta function:
\begin{align}\label{diff-delta}
\bar{\delta}^{(D)}(x) \equiv \ell^{D} \delta^{(D)}(x).
\end{align}
This definition can be used to write down a fully invariant definition of the functional determinant $ \overline{\operatorname{det}}$ as well; this is instrumental in ensuring that the functional measure is invariant as well.

Using these frame-invariant ingredients, we may define the grand configuration-space metric for the general theory \eqref{1.1} to be
\begin{equation}
\mathcal{G}_{i j} 
=\ell^{2}\left(\begin{array}{cc}
f P_{\mu \nu \rho \sigma} & -\frac{3}{4} f_{, B} g_{\mu \nu} \\
-\frac{3}{4} f_{, A} g_{\rho \sigma} & k_{A B}
\end{array}\right) \bar{\delta}^{(D)}\left(x_{I}-x_{J}\right)
\end{equation}
where $\bar{\delta}^{(D)}\left(x_{I}-x_{J}\right) \equiv \delta^{(D)}\left(x_{I}-x_{J}\right) / \sqrt{-\bar{g}}\:$ is frame invariant. We can also ensure that the path integral volume element $  V_F [\overline{\mathcal{D}} \Phi] \sqrt{\overline{\operatorname{det}}\left(\mathcal{G}_{i j}\right)}$ is invariant, as desired. To this end, one includes
the gauge fixing $\chi^{\mu}(\boldsymbol{\Phi})$ in a reparametrization invariant manner, as well as the Faddeev-Poppov determinant \cite{faddeev}
\begin{equation}\label{fp-determinant}
    V_{\mathrm{FP}}=\overline{\operatorname{det}}\left(\frac{\bar{\delta} \chi^{\mu}(\boldsymbol{x})}{\bar{\delta} \xi^{\nu}(\boldsymbol{y})}\right)
\end{equation}
in the path integral measure, with $\xi^\mu$ being the gauge parameters,



The diffeomorphism and frame invariant effective action then reads
\begin{equation}\label{invariant_effective_action}
\exp \left(\frac{i}{\hbar} \Gamma[\vvarphi]\right)=\int[\overline{\mathcal{D}} \boldsymbol{\Phi}] \mathcal{M}[\boldsymbol{\Phi}] \exp \left\{\frac{i}{\hbar}\left[S[\boldsymbol{\Phi}]+\frac{\bar{\delta} \Gamma[\boldsymbol{\varphi}]}{\bar{\delta} \varphi^{i}} \Sigma^{i}[\boldsymbol{\varphi}, \boldsymbol{\Phi}]\right]\right\},
\end{equation}
where $\boldsymbol{\varphi}=\left(g^{\mu \nu}, \phi\right)$ and the modified functional derivative has been defined using the frame-invariant delta function \eqref{diff-delta}.
The path integral measure is in \eqref{invariant_effective_action} is given by
\begin{equation}
 \quad \mathcal{M}[\boldsymbol{\Phi}]=V_{\mathrm{FP}} \sqrt{\overline{\operatorname{det}}\left(\mathcal{G}_{i j}\right)}.
\end{equation}
Equation \eqref{invariant_effective_action} is the standard VDW expression for the covariant effective action where now the integral measure and functional derivatives have been written in a frame invariant way using \eqref{diff-delta} and \eqref{fp-determinant}. 

Note that the configuration space vector $\Sigma^a\left[\boldsymbol{\varphi}, \boldsymbol{\Phi}\right]$ that has replaced the non-covariant term $(\varphi^a- \phi^a)$ is related to the geodesic tangent vector $\sigma^{a}\left[\boldsymbol{\varphi}, \boldsymbol{\Phi}\right]$ by \cite{dewitt1}
\begin{equation}\label{Sigma_boson}
    \Sigma^{a}\left[\boldsymbol{\varphi}, \boldsymbol{\Phi}\right]=\left(C^{-1}[\boldsymbol{\varphi}]\right)^{a}\:_{b}\:\: \sigma^{b}\left[\boldsymbol{\varphi}, \boldsymbol{\Phi}\right]
\end{equation}
where the coefficient $C_{\:\:b}^{a}[\boldsymbol{\varphi}]= \left\langle \nabla_{b}\sigma^{a}\left[\boldsymbol{\varphi}, \boldsymbol{\Phi}\right]\right\rangle_\Sigma$ ensures that tadpoles evaluate to zero.
The effective action defined in this way satisfies the important property \cite{finn}
\begin{equation}
    \Gamma\left[\boldsymbol{\varphi} ; \ell(\phi), f(\phi), k_{A B}(\phi), V(\phi)\right]=\Gamma\left[\tilde{\varphi}(\varphi) ; \tilde{\ell}(\phi), \tilde{f}(\phi), \tilde{k}_{A B}(\phi), \tilde{V}(\phi)\right],
\end{equation}
where the transformations of the field and model functions are given by \eqref{1.6}, \eqref{7.3} and \eqref{8.2}. Thus, by construction, $\Gamma$ is manifestly frame invariant. In particular, the functional form of $\Gamma$ does not change under transformations of the fields and model functions. \newline

We can expand the effective action perturbatively as $\Gamma[\boldsymbol{\varphi}] = \sum_{l=0} \hbar^l \Gamma^{(l)}[\boldsymbol{\varphi}]$ and compute it explicitly order by order using the background field method \cite{abbott, alvarez}. Expanding the action covariantly as
\begin{equation}
    S\left[\boldsymbol{\Phi}+ \boldsymbol{\varphi}\right]=S[\boldsymbol{\varphi}]+\sum_{N=1} \frac{1}{N !} S_{a_1 \ldots a_N}[\boldsymbol{\varphi}] \Phi^{a_1} \ldots \Phi^{a_N},
\end{equation}
where
\begin{equation}
    S_{a_1\ldots a_N}[\boldsymbol{\varphi}] \equiv \left. {\nabla}_{a_1}\ldots {\nabla}_{a_N}S[\boldsymbol{\Phi}]\right|_{\boldsymbol{\Phi}\:=\: \boldsymbol{\varphi}},
\end{equation}
we can find the one- and two-loop covariant effective action to be
\begin{align}\label{1-loops}
\Gamma^{(1)}[\varphi]=& \frac{i}{2} \ln \overline{\operatorname{det}}\left(\nabla^{a} \nabla_{b} S\right), \\
\Gamma^{(2)}[\varphi]=&-\frac{1}{8} \Delta^{a b} \Delta^{c d} \nabla_{(a} \nabla_{b} \nabla_{c} \nabla_{d)} S +\frac{1}{12} \Delta^{a b} \Delta^{c d} \Delta^{e f}\left(\nabla_{(a} \nabla_{c} \nabla_{e)} S\right)\left(\nabla_{(b} \nabla_{d} \nabla_{f)} S\right),\label{2-loops}
\end{align}
where the parentheses imply symmetrisation over the indices enclosed, $\Delta_{a b} \equiv\nabla_{a} \nabla_{b} S$ and its inverse, defined by virtue of $\Delta^{ab}\Delta_{bc}=\delta^{a}_{\:\:c}$ with $\Delta^{ab}=\Delta^{ba}$, is the propagator. Note that the expressions for the one- and two-loop effective action in \eqref{1-loops}  and \eqref{2-loops} are consistent with \cite{ellicot}.

This concludes our treatment of bosonic fields. However, defining a field-space for fermions and equipping it with a unique metric has been an outstanding problem. This is going to be the focus of the following sections. 

\section{A Short Overview of Supermanifolds}\label{super}

To extend the covariant formulation of QFT to theories with fermionic degrees of freedom we will employ the mathematical language of Supermanifolds. Supermanifolds are the natural extension of a Riemannian manifold when Grassman valued coordinates are included \cite{berenzin1, berenzin2, kostant, batchelor1, batchelor2, leites} and their construction follows closely the standard prescription of differential geometry. While the topic is vast and mathematically intricate \cite{dewitt, rogers}, in this section we limit ourselves to review only those basic properties that play a direct role in the construction of a fermionic field space. \newline

Let us start setting the notation by considering a supermanifold with $n$ commuting coordinates and $m$ anti-commuting coordinates. General coordinates in the supermanifold chart are denoted with the use of Greek letters superscripts as $x^{\alpha}$, $\alpha = (1,2, \cdots, n+m)$. When performing a diffeomorphism in the presence of anti-commuting coordinates
\begin{equation}\label{3.2}
    x^\a \rightarrow \tilde{x}^\a = \tilde{x}^\a(\boldsymbol{x}),
\end{equation}
one needs to distinguish between left and right differentiation. The two operations are related by
\begin{equation}\label{3.1}
\frac{\overrightarrow{\partial}}{\partial x^\a} \tilde{x}^\b=(-1)^{\alpha(\b+1)} \tilde{x}^\b \frac{\overleftarrow{\partial}}{\partial x^\a}.
\end{equation}
where we have introduced a notation convention that will be employed for the rest of the paper. 
Each symbol in an exponent of ($-1$) is not meant to be taken literally but as a label taking the value of~$1$ for anticommuting quantities and $0$ for commuting quantities. 
Hence, \eqref{3.1} tells us that right and left derivative differ only when differentiating a commuting object with respect to a anticommuting coordinate. In line with this new convention, indices can be contracted in a straightforward way only when adjacent and factors of $(-1)$ have to be introduced otherwise and every time the position of two indices is switched, e.g. $x^\a\:{}_\b x = (-1)^{\a \b}{}_\b x\: x^\a.$
The expressions in \eqref{3.1} are defined to be the left and right Jacobians:
\begin{equation}
    _\a J^\b = \frac{\overrightarrow{\partial}}{\partial x^\a}\tilde{x}^\b \equiv\: \overrightarrow{_\a\partial}\:\tilde{x}^\b,\quad \qquad \quad ^\b J^{\st}_\a=\:^\b \tilde{x} \frac{\overleftarrow{\partial}}{\partial x^\a}\equiv\:^\b\tilde{x}\: \overleftarrow{\partial_\a}.
\end{equation}
Accordingly, we introduce four types of tensors which transform with either left, right Jacobian or their inverse:
\begin{equation}\label{3.3}
    \begin{aligned}
^\a \tilde{X} &=\:^\a J^{\st}_\b\:^\b X,\qquad \qquad \quad \hspace{1mm}
\tilde{X}^\a =\: X^\b\: _\b J^\a,\\
_\a\tilde{X} &=\:_\a (J^{-1})^\b\:_\b X,\qquad \qquad
\tilde{X}_\a =\: X_\b\:^\b(J^{-1})^{\st}_\a.
    \end{aligned}
\end{equation}
The superscript $\st$ stands for \textit{supertransposition}. This operation is defined differently for each possible index placement:
\begin{equation}\label{3.4}
    \begin{aligned}
_\alpha M ^{\beta \:\st} &= (-1)^{\alpha(\beta +1)}\enspace ^\beta M _\alpha, \qquad \quad \hspace{6mm}
  { }^{\alpha} M_{\beta}^{\: \st}=(-1)^{\beta(\alpha+1)}\enspace_{\beta} M^{\alpha},\\
  { }_{\alpha} M_{\beta}^ {\st}&=(-1)^{\alpha+\beta+\alpha \beta} \enspace_{\beta} M_{\alpha},\qquad \quad 
  ^{\alpha} M^{\beta \:\st}=(-1)^{\alpha \beta}\enspace ^\beta M^{\alpha},
\end{aligned}
\end{equation}
leading to four different types of \textit{supermatrices}.
For a supermatrix with both indices in the upper position, $ ^{\alpha} M^{\beta}$,  or with both indices in the lower position, $ _{\alpha} M_{\beta}$, we introduce the notion of sypersymmetricity if $ M^{\st}= M $ and anti-sypersymmetricity if $M^{\st}= -M$.\newline

For supermatrices with one index in the upper position and one in the lower position, i.e. $ ^{\alpha} M_{\beta}$ or $_{\alpha} M^{\beta}$, we can also introduce the operation of \textit{supertrace}:
\begin{equation}\label{1.13}
    \operatorname{str}\:(_\a M ^\b)= (-1)^\a\:_\a M^\a= M_\a\:^\a, \qquad \qquad  \operatorname{str}\:(^\a M _\b)= (-1)^\a\:^\a M_\a=(-1)^\a\: M^\a\:_\a.
\end{equation}
The operation of supertrace produces \textit{superscalars}, i.e. quantities invariant under the transformations listed in \eqref{3.3} and \eqref{3.4}. Notice that $\operatorname{str}\:(_\a M ^\b)= \operatorname{str}\:(^\a M _\b)$ due to the cyclicity of the supertrace. 


The last operation to introduce is that of \textit{superdeterminant} \cite{berenzin2} which is defined for each type of supermatrix in \eqref{3.4}. A general supermatrix has the form 
\begin{equation}\label{matrix}
    { }_{\alpha} M_{\beta}=\left(\begin{array}{cc}
{ }_{A} A_{B} & { }_{A} C_{J} \\
{ }_{I} D_{B} & { }_{I} B_{J}
\end{array}\right),
\end{equation}
where ${ }_{A} A_{B}$ and ${ }_{I} B_{J}$ are $n \times n$ and $m \times m$ matrices of commuting numbers, respectively and ${ }_{A} C_{J}$ and ${ }_{I} D_{B}$ are $n \times m$ and $m \times n$ matrices of anticommuting numbers, respectively. The superdeterminant of \eqref{matrix} is given by
\begin{equation}
    \operatorname{sdet} M=\frac{\operatorname{det}\left(A-C B^{-1} D\right)}{\operatorname{det} B}= \frac{\det A}{\det (B - C A^{-1}D)}.
\end{equation}
Taking the superdeterminant of a supermatrix allows one to write an integral measure that is invariant under diffeomorphisms \eqref{3.2} of the supermanifold  \cite{berenzin2}:
\begin{equation}
    \sqrt{\left|\operatorname{sdet}(M)\right|} \d^{n+m} x.
\end{equation}
Lastly, a supermanifold is Riemannian if endowed with a real rank-2 tensor field $_\a G_\b$ known as the \textit{metric}. The supermanifold metric has to be supersymmetric and non-singular \cite{dewitt}. The inverse metric $^\a G^\b$, defined from the identity $ _\a G_\c \:G^{\c \b}=\: _\a\delta^{\b}$, satisfies
\begin{equation}
    ^\a G^\b = G^{\a \b}= (-1)^{\a \b}\: G^{\b \a} .
\end{equation}
Making use of the metric, we can write the general line element of the supermanifold as
\begin{equation}
    \mathcal{D}\Sigma^2 = \d x^\a\: _\a G_\b (x)\: \d x^\b.
\end{equation}

\section{Living on a Supermanifold: Field Space for Scalar-Fermionic Theories}\label{section5}
The scalar-fermion field space is built with $N$ real scalar fields and $M$ Dirac fermions yielding $8M$ anticommuting coordinates in 4-dimensional spacetime. General coordinates in the field-space supermanifold are
\begin{equation}
  \boldsymbol{\Phi} \equiv \left\{\Phi^{\alpha}\right\}=\left(\phi^{A}\:,\: \psi_{a}^{1}\:,\: \bar{\psi}_{\dot{a}}^{1}\:,\: \psi_{a}^{2}\:,\: \bar{\psi}_{\dot{a}}^{2}\:,\: \ldots\right),
\end{equation}
where $a$ and $\dot{a}$ are spinor indices. In what follows, capital Latin letters from the beginning of the alphabet denote bosonic degrees of freedom and capital letters from the end of the alphabet will denote fermionic ones.
The general frame covariant Lagrangian of a scalar theory with fermions can be written as \cite{finn2}
\begin{equation}\label{lagrangian}
    \mathcal{L}=\frac{1}{2} g^{\mu \nu} \partial_{\mu} \Phi^{\alpha}\:_{\alpha} k_{\beta}(\boldsymbol{\Phi})\: \partial_{\nu} \Phi^{\beta}+\frac{i}{2} \zeta_{\alpha}^{\mu}(\boldsymbol{\Phi})\: \partial_{\mu} \Phi^{\alpha}-U(\boldsymbol{\Phi}).
\end{equation}
The model functions $_{\alpha} k_{\beta}(\boldsymbol{\Phi})$, $\zeta_{\alpha}^{\mu}(\boldsymbol{\Phi})$, $U(\boldsymbol{\Phi})$ are respectively a rank-2 field-space tensor (with ${ }_{\alpha} k_{I}={ }_{I} k_{\alpha}=0$), a mixed spacetime and field-space vector, a scalar describing the potential and Yukawa sector. Note that in the absence of fermions $_{\alpha} k_{\beta}$  would be the bosonic field-space metric \cite{finn}. The function $\zeta_{\alpha}^{\mu}$ appears exclusively in fermionic theories since the Lagrangian for a scalar theory cannot contain a single derivative term.

Given a Lagrangian, the model functions can be extracted according to the following prescription
\begin{equation}\label{4.7}
    \begin{aligned}
{ }_{\alpha} k_{\beta} &=\frac{g_{\mu \nu}}{4} \frac{\vec{\partial}}{\partial\left(\partial_{\mu} \Phi^{\alpha}\right)} \mathcal{L} \frac{\overleftarrow{\partial}}{\partial\left(\partial_{\nu} \Phi^{\beta}\right)}, \\
\zeta_{\alpha}^{\mu} &=\frac{2}{i}\left(\mathcal{L}-\frac{1}{2} g^{\mu \nu} \partial_{\mu} \Phi_{\alpha}^{\alpha} k_{\beta} \partial_{\nu} \Phi^{\beta}\right) \frac{\overleftarrow{\partial}}{\partial\left(\partial_{\mu} \Phi^{\alpha}\right)}. \\
\end{aligned}
\end{equation}
In the field-space supermanifold, general reparametrizations of the fields 
\begin{equation}\label{field-rep}
    \Phi^{\alpha} \rightarrow \widetilde{\Phi}^{\alpha}=\widetilde{\Phi}^{\alpha}(\Phi)
\end{equation}
correspond to diffeomorphisms and hence invariance under \eqref{field-rep} can be achieved applying differential supergeometry techniques as outlined in Section \ref{super}.

\subsection{Deriving the Grand Metric}
To complete the construction of the field-space supermanifold and equip with a metric, we need to construct a pure field-space covector from $\zeta_{\alpha}^{\mu}$. Since by construction this tensor cannot depend on derivatives of the fields and there are no spacetime covectors in \eqref{lagrangian}, the spacetime index $\mu$ in $\zeta_{\alpha}^{\mu}$ must arise from a $\gamma^{\mu}$ matrix. Hence, we can construct a pure field-space vector by introducing a new type of differentiation with respect to the $\gamma^{\mu}$ matrix:
\begin{equation}\label{5.4}
    \zeta_{\alpha}=\frac{1}{4} \frac{\delta \zeta_{\alpha}^{\mu}}{\delta \gamma^{\mu}}.
\end{equation}
In analogy with how the field strength tensor $F_{\mu \nu}$ is constructed in QED, we derive the following rank-2 field-space tensor \cite{finn2}
\begin{equation}\label{2.12}
    { }_{\alpha} \lambda_{\beta}=\frac{1}{2}\left(\frac{\overrightarrow{\partial}}{\partial \Phi^{\alpha}} \zeta_{\beta}-(-1)^{\alpha+\beta+\alpha \beta} \frac{\overrightarrow{\partial}}{\partial \Phi^{\beta}} \zeta_{\alpha}\right).
\end{equation}
By construction \eqref{2.12}, $\:_{\alpha} \lambda_{\beta}$ is anti-supersymmetric and singualar in the presence of scalar fields. Nonetheless one can combine $\:_{\alpha} \lambda_{\beta}$ and $\:_{\alpha} k_{\beta}$ in such a way as to generate a matrix that is non-singular:
\begin{equation}\label{2.14}
  _{\alpha} \Lambda_{\beta}=\:  _{\alpha} k_{\beta} \:+\:  _{\alpha} \lambda_{\beta}.
\end{equation}
This matrix will play a fundamental role in the definition of the field-space metric. 

\subsection{Free Theory Example}
To understand how to extract and utilise the model functions, consider the following canonically normalised free theory example
\begin{equation}\label{4.5}
    \begin{aligned}
\mathcal{L}=& \sum_{A \in \text { scalars }}\left[\frac{1}{2} g^{\mu \nu} \partial_{\mu} \phi^{A} \partial_{\nu} \phi^{A}-\frac{1}{2} m_{A}^{2}\left(\phi^{A}\right)^{2}\right] \\
&+\sum_{X \in \text { fermions }}\left[\frac{i}{2}\left(\bar{\psi}^{X} \gamma^{\mu} \partial_{\mu} \psi^{X}-\partial_{\mu} \bar{\psi}^{X} \gamma^{\mu} \psi^{X}\right)-m_{X} \bar{\psi}^{X} \psi^{X}\right].
\end{aligned}
\end{equation}
Using \eqref{4.7} we can read off the model functions to be
\begin{align}
{ }_{\alpha} k_{\beta} &=\left(\begin{array}{cc}
\delta_{A B} & \mathbf{0}_{N \times 8 M} \\
\mathbf{0}_{8 M \times N} & \mathbf{0}_{8 M \times 8 M}
\end{array}\right), \\
\zeta_{\alpha}^{\mu}\label{zeta} &=\left(\mathbf{0}_{N}, \bar{\psi}_{\dot{a}}^{1} \gamma_{a a}^{\mu}, \gamma_{\dot{a} a}^{\mu} \psi_{a}^{1}, \bar{\psi}_{\dot{b}}^{2} \gamma_{\dot{b} b}^{\mu}, \gamma_{\dot{b} b}^{\mu} \psi_{b}^{2}, \ldots\right). 
\end{align}
Differentiating \eqref{zeta} with respect to $\gamma^{\mu}$ according to \eqref{5.4} and substituting this into \eqref{2.14} 
we can compute the matrix $\:_{\alpha} \Lambda_{\beta}$ to be
\begin{equation}\label{2.20}
    { }_{\alpha} \Lambda_{\beta}={ }_{\alpha} N_{\beta} \equiv\left(\begin{array}{c|ccccc}
1_{N} & 0 & 0 & 0 & 0 & \ldots \\
\hline
0 & 0 & 1_{4} & 0 & 0 & \ldots \\
0 & 1_{4} & 0 & 0 & 0 & \ldots \\
0 & 0 & 0 & 0 & 1_{4} & \ldots \\
0 & 0 & 0 & 1_{4} & 0 & \ldots \\
\vdots & \vdots & \vdots & \vdots & \vdots & \ddots
\end{array}\right) .
\end{equation}

\subsection{Properties of the Grand Field-Space Metric}
The Grand metric $_\a G_\b (\boldsymbol{\Phi})$ for the field-space supermanifold should satisfy the following requirements:
\begin{enumerate}
    \item The metric should be determined solely and uniquely from the action;
    \item The metric should be a supersymmetric rank-2 field-space tensor;
    \item The metric should not be singular in order to produce a non-zero line element;
    \item The metric should be ultralocal, i.e. it should not depend on derivatives of the fields;
    \item The metric should have the local form on each point of the field-space supermanifold \cite{dewitt, finn2}:
    \begin{equation}    \label{6.1}
{ }_{a} H_{b} \equiv\left(\begin{array}{cccccc}
1_{N} & 0 & 0 & 0 & 0 & \cdots \\
0 & 0 & 1_{4} & 0 & 0 & \cdots \\
0 & -1_{4} & 0 & 0 & 0 & \cdots \\
0 & 0 & 0 & 0 & 1_{4} & \cdots \\
0 & 0 & 0 & -1_{4} & 0 & \cdots \\
\vdots & \vdots & \vdots & \vdots & \vdots & \ddots
\end{array}\right).
\end{equation}
\end{enumerate}
The matrix  $_\a \Lambda _b$ is non-singular but it is not supersymmetric and therefore cannot be used as the field-space metric. To find a suitable metric for the field-space that satisfies the requirements listed above we will make use of the vielbein formalism \cite{schwinger}. 
In a local inertial frame we expect the fields to be locally canonical \cite{riemann} with kinetic terms as in \eqref{4.5} and $_a \Lambda_b =\: _a N_b$. Thus we can write
\begin{equation}\label{6.3}
     _\a \Lambda _\b \:=\: _\a e^a\:\:_a N_b\:\:^be^{\st}_\b,
\end{equation}
where $_\a e^a$ are the vielbeins and $\:_a N_b$ is as defined in \eqref{2.20}. 
Once the vielbeins are found from \eqref{6.3}, the field-space metric $ _\a G _\b$ can be obtained as \cite{finn2}
\begin{equation}\label{6.2}
    _\a G _\b \:=\: _\a e^a\:\:_a H_b\:\:^be^{\st}_\b,
\end{equation}
where $_a H_b$ is the local metric defined in \eqref{6.1}. 

Once the metric is determined from \eqref{6.2}, one can then define the Christoffel symbols as
\begin{equation}\label{6.9}
 \Gamma ^\a_{\:\: \b \c} = \frac{1}{2}\:^\a G^{ \delta}\left[\:_\delta G_{ \b}\:\overleftarrow{\partial_\c}+(-1)^{\b \c}\:_\delta G_{\c}\:\overleftarrow{\partial_\b}-(-1)^{\b}\:\overrightarrow{_\delta\partial} \:_\b G_{\c}\right],
\end{equation}
and from this covariant derivatives on the field-space
\begin{equation}
\begin{aligned}
X^{\a}\overleftarrow{\nabla}_{\b} &=X^\a \overleftarrow{\partial_\b} + \Gamma_{\:\:\b \c}^{\a}X^\c,\\
X_{\a}\overleftarrow{\nabla}_{\b} &=X_{\a\:}\overleftarrow{\partial_{\b}}-X_{\c} \Gamma_{\:\:\a \b}^{\c}.
\end{aligned}
\end{equation}
The field-space Riemann tensor is then defined as 
\begin{equation}\label{6.10}
R^\a_{\:\:\b \c \delta}= \left[-\Gamma^{\a}_{\:\:\b \c}\:\overleftarrow{\partial_{\delta}}+(-1)^{\c \delta}\: \Gamma^{\a}_{\:\:\b \delta}\:\overleftarrow{\partial_{\c}}+(-1)^{\c(\sigma+\b)}\: \Gamma_{\:\:\sigma \c}^{\a} \Gamma_{\:\:\b \delta}^{\sigma}\right.\left.-(-1)^{\delta(\sigma+\b+\c)}\: \Gamma_{\:\:\sigma \delta}^{\a} \Gamma_{\:\:\b \c}^{\sigma}\right].
\end{equation}
Promoting the field-space to a configuration-space is now straightforward. Coordinates in configuration-space are denoted by
\begin{equation}\label{6.11}
    \Phi^{\widehat{\alpha}} \equiv \Phi^{\alpha}\left(\boldsymbol{x}_{\alpha}\right),
\end{equation}
and the definition of the metric and metric connections are generalised as
\begin{equation}
    { }_{\widehat{\alpha}} G_{\widehat{\beta}}={ }_{\alpha} G_{\beta}\: \delta^{(4)}\left(\boldsymbol{x}_{\alpha}-\boldsymbol{x}_{\beta}\right),\label{csmetric}
\end{equation}
and 
\begin{equation}
    { }^{\widehat{\alpha}} \Gamma_{\widehat{\beta} \widehat{\gamma}}={ }^{\alpha} \Gamma_{\beta \gamma}\: \delta^{(4)}\left(\boldsymbol{x}_{\alpha}-\boldsymbol{x}_{\beta}\right) \delta^{(4)}\left(\boldsymbol{x}_{\alpha}-\boldsymbol{x}_{\c}\right).
\end{equation}
Making use of the metric \eqref{csmetric}, the configuration space line element is defined as
\begin{equation}\label{line}
    \mathcal{D}\Sigma^2 = \d \Phi^{\widehat{\alpha}} \:\:{ }_{\widehat{\alpha}} G_{\widehat{\beta}} (\boldsymbol{\Phi})\:\d \Phi^{\widehat{\b}}.
\end{equation}
Notice that \eqref{line} is written in terms of fully contracted configuration space tensors and is therefore a superscalar.

Finally, the reparametrization invariant integral measure needed to construct a covariant path integral reads
\begin{equation}
    [\mathcal{D} \mathcal{M}]=\sqrt{|\operatorname{sdet} G|}\left[\mathcal{D}^{N+8 M} \boldsymbol{\Phi}_q\right],
\end{equation}
where now $\boldsymbol{\Phi}_q$ denotes collectively the $\Phi^{\widehat{\a}}$ coordinates on the configuration-space as defined through \eqref{6.11}.

\section{Grand Covariant Effective Action with Fermions}\label{section6}
We now possess all the tools needed to construct a reparametrization invariant expression for the effective action using the VDW formalism \cite{vilk1, vilk2, dewitt1,rebhan, ellicot}. The implicit equation for the effective action reads \cite{finn2}
\begin{equation}\label{7.1}
  \exp \left(\frac{i}{\hbar} \Gamma[\boldsymbol{\Phi}]\right)=\int \sqrt{|\operatorname{sdet} G|}\left[\mathcal{D} \boldsymbol{\Phi}_q\right] \exp \left(\frac{i}{\hbar} S\left[\boldsymbol{\Phi}_q\right]+\frac{i}{\hbar} \int \d^{4} x \sqrt{-g}\: \Gamma[\boldsymbol{\Phi}] \frac{\overleftarrow{\partial}}{\partial \Phi^{\alpha}} \Sigma^{\alpha}\left[\boldsymbol{\Phi}, \boldsymbol{\Phi}_q\right]\right) , 
\end{equation}
where $\boldsymbol{\Phi}$ denotes the mean field and $\Sigma^{\alpha}\left[\boldsymbol{\Phi}, \boldsymbol{\Phi}_q\right]$ is defined in analogy to \eqref{Sigma_boson} to be a superposition of supergeodesic tangent vectors $\sigma^{\alpha}\left[\boldsymbol{\Phi}, \boldsymbol{\Phi}_q\right]$:
\begin{equation}
    \Sigma^{\alpha}\left[\boldsymbol{\Phi}, \boldsymbol{\Phi}_q\right]=\left(C^{-1}[\boldsymbol{\Phi}]\right)^{\alpha}{ }_{\beta}\:\: \sigma^{\beta}\left[\boldsymbol{\Phi}, \boldsymbol{\Phi}_q\right],
\end{equation}
where the coefficient now reads $C_{\:\:\beta}^{\alpha}[\boldsymbol{\Phi}]= \left\langle\sigma^{\alpha}\left[\boldsymbol{\Phi}, \boldsymbol{\Phi}_q\right] \overleftarrow{\nabla}_{\beta}\right\rangle_\Sigma$. 
Using a covariant expansion for the action, the expressions for the one- and two-loop covariant effective action are found to be
\begin{align}
   \Gamma^{(1)}[\boldsymbol{\Phi}]&= \frac{i}{2}\ln \operatorname{sdet}\:^{\widehat{\a}} S_{\widehat{\b}}= \frac{i}{2} \operatorname{str}\ln \:^{\widehat{\a}} S_{\widehat{\b}} \:,\label{7.5}\\
    \Gamma^{(2)}[\boldsymbol{\Phi}]
    &=-\frac{1}{8}S_{\{\widehat{\a} \widehat{\b} \widehat{\c} \widehat{\delta}\}}\Delta^{\delta \widehat{\c}}\Delta^{\widehat{\b}\widehat{\a}}+\frac{1}{12}(-1)^{\widehat{\b}\widehat{\c} + \widehat{\epsilon}(\widehat{\b} + \widehat{\delta})}\:S_{\{\widehat{\epsilon} \widehat{\c} \widehat{\a}\}}\Delta^{\widehat{\a} \widehat{\b}}\Delta^{\widehat{\c}\delta}\Delta^{\widehat{\epsilon} \widehat{\zeta}}{}_{\{ \widehat{\zeta} \widehat{\delta} \widehat{\b}\}}S\:, \label{7.6}
\end{align}
where $^{\widehat{\a}} S_{\widehat{\b}}= \:^{\widehat{\a}} \overrightarrow{\nabla}S\overleftarrow{\nabla}_{\widehat{\b}}$ and $\:_{\widehat{\a}} S_{\widehat{\b}}=\:_{\widehat{\a}} \overrightarrow{\nabla}S\overleftarrow{\nabla}_{\widehat{\b}}=\:_{\widehat{\a}}( \Delta^{-1})_{\widehat{\b}}$. The inverse $\Delta^{{\widehat{\a}} {\widehat{\b}}}$, defined through 
$\Delta^{\widehat{\a}\widehat{\c}}\:_{\widehat{\c}}S_{\widehat{\b}}=\:^{\widehat{\a}}\delta _{\widehat{\b}}$\:, is the frame-covariant propagator. 
It is easy to show that the propagator and its inverse are both supersymmetric. The notation $\{\ldots\}$ denotes the operation of supersymmetrization of the indices enclosed \cite{finn2}:
\begin{equation}
    \{\a_1 \ldots \a_2\}=\frac{1}{n!}\sum_{P}(-1)^P P[\a_1 \ldots \a_2],
\end{equation}
where $P$ spans over all possible permutations of the $n$ indices and the factor $(-1)^P$ yields $-1$ for odd number of swaps between fermionic indices and $0$ otherwise. 
Note that both \eqref{7.5} and \eqref{7.6} are superscalars as one would expect. The presence of the pre-factor  $(-1)$ in the second term of \eqref{7.6} does not spoil covariance but is consistent with the convention that only adjacent pairs of indices can be contracted straightforwardly as explained in Section \ref{super}. The one- and two-loop effective action results in \eqref{7.5} and \eqref{7.6} reduce to \eqref{1-loops} and \eqref{2-loops} in the absence of fermionic variables.

\section{Conclusions}\label{section7}
We have developed a fully covariant formalism for scalar-tensor theories of gravity by modifying the standard Vilkovisky-DeWitt effective action to include a integral measure and functional derivatives that are both frame and diffeomorphisms invariant. We have constructed the field-space supermanifold for scalar-fermion theories and developed a rigorous algorithm for calculating the field-space metric from the classical action by means of vielbeins. We have extended the VDW formalism on Supermanifolds and obtained an expression for the quantum effective action for theories with fermions that is both frame- and reparametrization-invariant. 

Since the VDW formalism has already been successfully applied to scalar theories, gauge theories and gravity, the inclusion of fermions was the last milestone for a complete geometrisation of a broad spectrum of QFTs. The natural next step would be to construct a field-space supermanifold for realistic theories of high energy physics such as the Standard Model. We postpone this task to future work.

\acknowledgments
KF and VG acknowledge support by the University of Manchester through the President's Doctoral Scholar Award. The work of AP is supported by the Lancaster–Manchester–Sheffield Consortium for Fundamental Physics, under STFC research
grant ST/T001038/1.

\newpage
\appendix
\section{Single Fermion Example}\label{appendix}
As an explicit example of computation of the field-space metric, consider the theory with one scalar field $\phi$ and one Dirac fermion $\psi$ described by the Lagrangian:
\begin{equation}
   \begin{aligned}
\mathcal{L}=& \frac{1}{2} k(\phi) \partial_{\mu} \phi \partial^{\mu} \phi-\frac{1}{2} h(\phi) \bar{\psi} \gamma^{\mu} \psi \partial_{\mu} \phi+\frac{i}{2} g(\phi) \bar{\psi} \gamma^{\mu} \partial_{\mu} \psi \\
&-\frac{i}{2} g(\phi) \partial_{\mu} \bar{\psi} \gamma^{\mu} \psi-Y(\phi) \bar{\psi} \psi-V(\phi)
\end{aligned}, 
\end{equation}
where $k, h, g, Y$ and $V$ are arbitrary real functions of $\phi$. Using \eqref{4.7}  to find the model functions and definitions \eqref{5.4} and \eqref{2.12}, we can construct the rank-2 tensor $_\a\Lambda_\b$ to be
\begin{equation}\label{2.6}  { }_{\alpha} \Lambda_{\beta}=\left(\begin{array}{ccc}
k & \frac{1}{2}\left(g^{\prime}-i h\right) \bar{\psi} & \frac{1}{2}\left(g^{\prime}+i h\right) \psi \\
\frac{1}{2}\left(g^{\prime}-i h\right) \bar{\psi} & 0 & g 1_{4} \\
\frac{1}{2}\left(g^{\prime}+i h\right) \psi & g 1_{4} & 0
\end{array}\right).
\end{equation}
Substituting this into \eqref{6.3}, we find the vielbeins to have the general form
\begin{equation}
    { }_{\alpha} e^{a}=\left(\begin{array}{ccc}
\sqrt{k} & \frac{g^{\prime}+i h}{2g}\psi\: x & \frac{g^{\prime}-i h}{2} \bar{\psi} x^{-1} \\
0 & x & 0 \\
0 & 0 & gx^{-1}
\end{array}\right),
\end{equation}
where $x$ is an arbitrary invertible symmetric  $4 \times 4$ matrix. The presence of the arbitrary matrix $x$ reflects that the vielbeins are not determined uniquely by the condition \eqref{6.3}. However, note that this ambiguity does not appear in the expression for the metric, because $x$ is eliminated when the products between the vielbein and its supertranspose is taken. 
Substituting the vielbeins into \eqref{6.2} 
we find the field-space metric to be
\begin{equation}\label{2.15}
{ }_{\alpha} G_{\beta}=\left(\begin{array}{ccc}
k-\frac{g^{\prime 2}+h^{2}}{2 g} \bar{\psi} \psi & -\frac{1}{2}\left(g^{\prime}-i h\right) \bar{\psi} & \frac{1}{2}\left(g^{\prime}+i h\right) \psi \\
\frac{1}{2}\left(g^{\prime}-i h\right) \bar{\psi} & 0 & g 1_{4} \\
-\frac{1}{2}\left(g^{\prime}+i h\right) \psi & -g 1_{4} & 0
\end{array}\right),
\end{equation}
which is supersymmetric as expected.
With the metric \eqref{2.15}, one can now calculate the metric connection components according to \eqref{6.9}. Using these to compute the Riemann tensor according to \eqref{6.10}, yields $R^\a_{\b \c \delta}=0$, i.e. the field-space for this theory is everywhere flat. This implies in turn that the theory \eqref{3.1} can be made canonical after a suitable reparametrization of the fields:
\begin{equation}\label{3.24}
    \phi \rightarrow \widetilde{\phi} = \int_{0}^{\phi}\sqrt{k(\phi^{\prime})}\d\phi^{\prime},\qquad \:\:
    \psi \rightarrow \widetilde{\psi} =\sqrt{g(\phi)}\exp\left(\frac{i}{2}\int_{0}^{\phi}\frac{h(\phi^{\prime})}{g(\phi^{\prime})}\d\phi^{\prime}\right)\psi,
\end{equation}
In this Cartesian frame the Lagrangian \eqref{3.1} now reads
\begin{equation}\label{3.25}
\mathcal{L}= \frac{1}{2} \partial_{\mu} \widetilde{\phi} \partial^{\mu} \widetilde{\phi}+\frac{i}{2} \widetilde{\bar{\psi}} \gamma^{\mu} \partial_{\mu} \widetilde{\psi} \\
-\frac{i}{2} \partial_{\mu} \widetilde{\bar{\psi}} \gamma^{\mu} \widetilde{\psi}-\widetilde{Y}(\widetilde{\phi}) \widetilde{\bar{\psi}} \widetilde{\psi}-\widetilde{V}(\widetilde{\phi}),
\end{equation}
with $\widetilde{Y}(\widetilde{\phi})=Y(\phi)/g(\phi)$ and $\widetilde{V}(\widetilde{\phi})= V(\phi)$.
For the canonically normalised theory \eqref{3.25} we can now compute the covariant propagator 
and from this the one-loop covariant effective action: 
\begin{equation}
\begin{aligned}
\Gamma[\Phi]=& S[\Phi]+\frac{i}{2} \operatorname{Tr} \ln \left\{\square+\widetilde{V}^{\prime \prime}(\widetilde{\phi})-\widetilde{\bar{\psi}}\left[2\widetilde{Y}^{\prime}(\widetilde{\phi})(-i\slashed{\partial} +\widetilde{Y}(\widetilde{\phi}))^{-1}\widetilde{Y}^{\prime}(\widetilde{\phi})-\widetilde{Y}^{\prime \prime}(\widetilde{\phi})\right] \widetilde{\psi}\right\} \\
&-i \operatorname{Tr} \ln (-i\slashed{\partial}+\widetilde{Y}(\widetilde{\phi})).
\end{aligned}
\end{equation}
This is consistent with previous results in the literature, e.g. see eq. (8.49) in \cite{zinn}.

\end{document}